 \documentclass[smallabstract,smallcaptions]{dccpaper}

\usepackage{amsmath,amssymb,amsfonts}
\usepackage{algorithmic}
\usepackage{graphicx}
\usepackage{textcomp}
\usepackage{subfig}
\usepackage{url}
\usepackage{xcolor}

\newlength{\figurewidth}
\newlength{\smallfigurewidth}

\graphicspath{{./images/}}

\begin{document}

\title
{\large
\textbf{Adaptive and  Scalable Compression of Multispectral Images using VVC}
}

\author{%
Philipp Seltsam, Priyanka Das, Mathias Wien\\[0.5em]
{\small\begin{minipage}{\linewidth}\begin{center}
\begin{tabular}{c}
    Institute of Imaging and Computer Vision \\
    RWTH Aachen University \\
    52056 Aachen, Germany \\
    \url{firstname.lastname@lfb.rwth-aachen.de}
\end{tabular}
\end{center}\end{minipage}
}}

\maketitle
\thispagestyle{empty}

\begin{abstract}
The VVC codec is applied to the task of multispectral image (MSI) compression using adaptive and scalable coding structures. In a ``plain'' VVC approach, concepts from picture-to-picture temporal prediction are employed for decorrelation along the MSI's spectral dimension. 
The popular principle component analysis (PCA) for spectral decorrelation is further evaluated in combination with VVC intra-coding for spatial decorrelation. 
This approach is referred to as PCA-VVC. A novel adaptive MSI compression algorithm, named HPCLS, is introduced, that uses PCA and inter-prediction for spectral and VVC intra-coding for spatial decorrelation. Further, a novel adaptive scalable approach is proposed, that provides a separately decodable spectrally scaled preview of the MSI in the compressed file. Information contained in the preview is exploited in order to reduce the overall file size. 
All schemes are evaluated on images from the ARAD HS data set containing outdoor scenes with a high variety in brightness and color. We found that ``Plain'' VVC is outperformed by both PCA-VVC and HPCLS. HPCLS shows advantageous rate-distortion (RD) behavior compared to PCA-VVC for reconstruction quality above 51\,dB PSNR. The performance of the scalable approach is compared to the combination of an independent RGB preview and one of HPCLS or PCA-VVC. The scalable approach shows significant benefit especially at higher preview qualities.
\end{abstract}

\section{Introduction}
Multispectral (MS) images capture a scene's spectro-spatial information in a set of 2D intensity images called spectral bands, where each band corresponds to a wavelength range within the electromagnetic spectrum. Due to it's remote sensing capability and non-intrusive character, MS imaging is used in numerous areas of application such as medicine, food processing and satellite imaging \cite{crane2011intraoperative}\cite{qin2013hyperspectral}\cite{hudak2007relationship}. The higher the spectral resolution, however, the higher the number of bands in the 3D MS image (MSI), which in turn generates large image files. This calls for MSI compression techniques, that, next to the spatial correlation, are able to efficiently exploit the high correlation along the spectral dimension, in order to lower transmission and storage costs.\\ 

MSI compression is an active field of research using a variety of techniques, which can be categorized into lossless and lossy compression. In lossless compression, only redundant information is discarded in order to reduce the amount of data. For lossy compression, also non-redundant information is discarded based on an importance criterion determined by the specific application. In the state of the art, most approaches use methods from transform coding, predictive coding, tensor decomposition and machine learning.

The linear Principle Component Analysis \cite{jolliffe2002principal} (PCA) is a popular choice among transforms used for spectral decorrelation. \cite{nagendran2020hyperspectral}-\nocite{du2009segmented,du2007hyperspectral,mei2018low}\cite{dragotti2000compression} use the PCA for spectral decorrelation in combination with a Discrete Wavelet Transform (DWT) for spatial decorrelation, which is also used for spatial compression in the JPEG2000 \cite{rabbani2002jpeg2000} standard. In \cite{tang2006three}, the DWT is performed on both spatial and spectral direction, resulting in a 3D DWT, and \cite{karami2012hyperspectral} uses a 3D Discrete Cosine Transform \cite{1672377} (DCT) followed by a support vector machine regression on the DCT coefficients for further compression. Several new codecs have been released since the launch of the JPEG2000 standard, which motivated Kwan et al.~\cite{kwan2019new} to test the performance of J2K, X264, X265, and Daala in combination with PCA on MSI compression.

Karami et al.~\cite{karami2012compression} combine a 2D DWT with a tensor decomposition of the 3D concatenation of wavelet coefficients, while in \cite{zhang2015compression} Zhang et al.~propose a pure tensor decomposition approach. 

In \cite{klimesh2005low}-\nocite{toivanen2005correlation}\nocite{zhang2007efficient}\nocite{li2019linear}\nocite{abrardo2011low}\cite{karaca2019superpixel}, linear DPCM prediction is used for compression of MSIs, where the predictor weights are usually determined by the least squares (LS) method. Intra-band prediction uses pixel values from within the same band for prediction, while inter-band prediction uses pixel values from previously coded bands to predict the current band. Toivanen et
al.~\cite{toivanen2005correlation} and Zhang and Liu \cite{zhang2007efficient} experiment with correlation-based band reordering in order to improve inter-band prediction performance. Li et al.~\cite{li2019linear} use correlation-based K-means clustering to obtain a set of reference bands to predict the entire MSI. In \cite{abrardo2011low}, Abrardo et al.~perform separate inter-band prediction on spatially non-overlapping rectangular image blocks, while Karaca and Güllü \cite{karaca2019superpixel} propose performing inter-band prediction on a superpixel basis, where superpixels are obtained using the simple linear iterative clustering algorithm \cite{achanta2010slic}.

Inter-band prediction using a non-linear multilayer propagation network of predetermined structure is proposed in \cite{dusselaar2019block}, where only the parameters of the network trained in the encoding stage have to be transmitted to the decoder side together with the prediction error. Deng et al.~in \cite{deng2020learning} employ an autoencoder network that directly learns a mapping from the MSI to a latent space representation of fewer degrees of freedom, that constitutes the compressed version of the MSI.\\

In this paper, Versatile Video Coding (VVC) \cite{Bross2021, VVC1}, the most recent video codec released by the JVET, is applied to the problem of MSI compression. In a ``plain'' VVC approach, VVC inter-prediction is employed for spectral decorrelation using a Group of Picture (GOP) structure has been adapted to treat the spectral dimension of the MSI image as the temporal dimension. This is used as the baseline to compare the other compression algorithms. VVC intra-coding is tested in combination with the popular PCA for spectral decorrelation, referred to as PCA-VVC. Further, two novel adaptive compression algorithms are introduced, that also use VVC intra-coding for spatial decorrelation. Hybrid Principle Component Least Squares (HPCLS) uses linear inter-band prediction to predict the entire MSI from a set of reference bands, that are obtained via PCA. The prediction error and the reference bands are further compressed via VVC intra-coding. A fourth compression algorithm is motivated by the desire of having a spectrally scaled preview of the MSI's contents included in the compressed MSI file.  The RGB Least Squares (RGBLS) algorithm provides such a preview in the form of a RGB representation of the MSI included in the compressed image file. The algorithm makes use of the information contained in the RGB base layer by predicting all MSI bands from the RGB values. Similar to HPCLS, the RGB and the prediction error image are coded using PCA and VVC. Experimental results obtained for the non-scalable PCA-VVC and HPCLS are used to asses the performance of the spectrally scaled solution.

The remainder of this paper is structured as follows: In section \ref{sec:dataset}, the data set used for evaluation is briefly introduced, followed by a description of the implemented compression algorithms in section \ref{sec:methods}. Section \ref{sec:experiments} features the experimental setup and evaluation results, which are then discussed in section \ref{sec:discussion}. Section \ref{sec:conclusion} draws a conclusion of the findings from this study and proposes possible future directions of research.

\section{Data Set}\label{sec:dataset}

\begin{figure}[htb]
	\centering
	\includegraphics[width=0.4\textwidth]{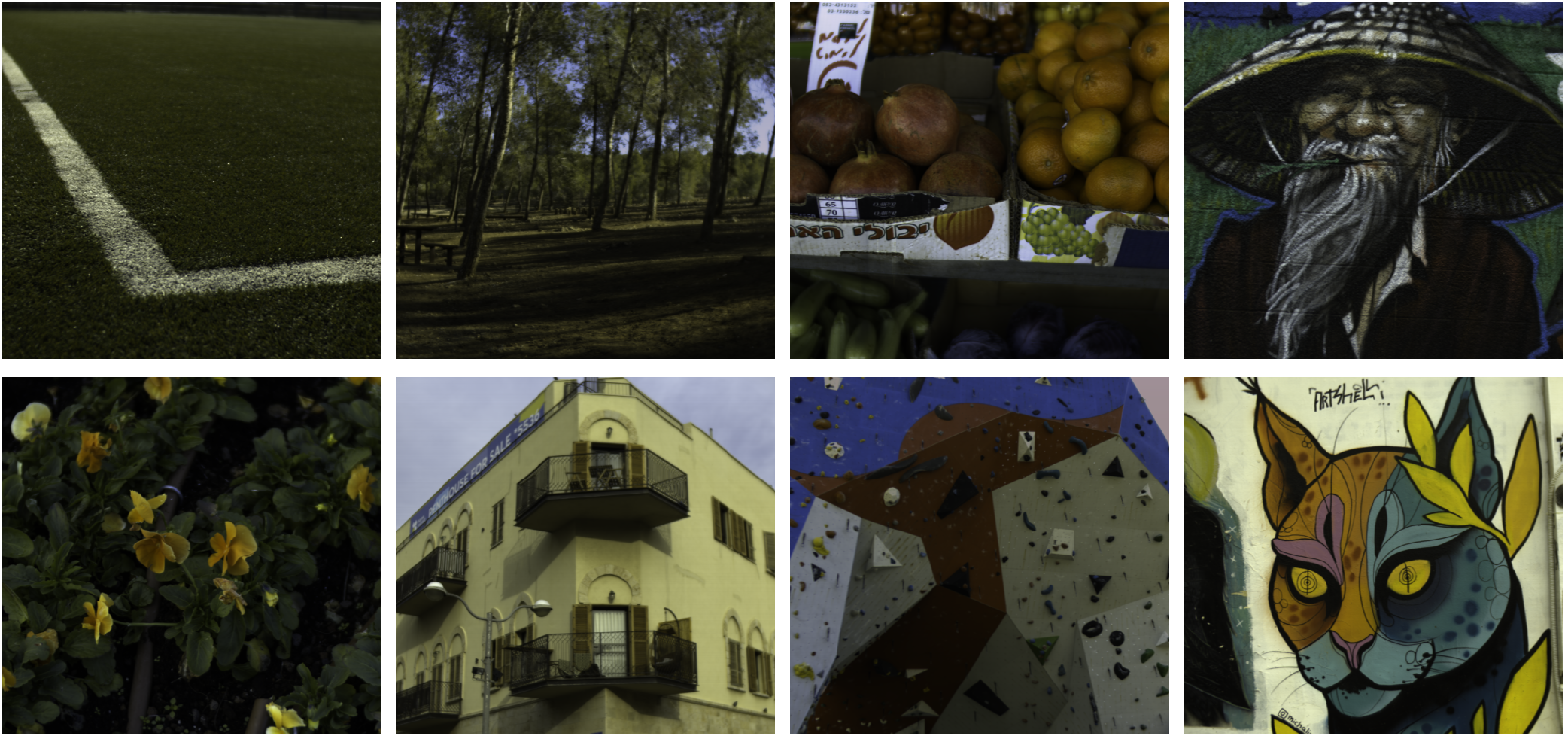}
	\caption[]{Example images from the ARAD HS data set}
	\label{fig:data}
\end{figure}

The ARAD HS data set published for the NTIRE2020 challenge on spectral reconstruction \cite{arad2020ntire} is used in this work. The data set comprises 470 images of outdoor scenes with a high variety in brightness and color. The 10 images that were used as validation data set in the challenge were selected for evaluation of the compression algorithms implemented in this paper. Exemplary RGB versions of images taken from the data set are shown in figure \ref{fig:data}. Each MSI comprises 31 \(482{\times} 512\) sized band images at 10\,nm increments in the visible 400-700\,nm range. Each image is cropped to yield four \(256{\times} 256{\times} 31\) sized images to fit to the VVC coding tree unit size of \(128 {\times} 128\), resulting in a total of 40 evaluation images.

\section{Methods}\label{sec:methods}

The compression algorithms introduced in the following are evaluated by their achieved reconstruction quality, measured in Peak Signal to Noise Ratio (PSNR), at a given compressed file size, measured in bits. The PSNR is defined as follows:

\begin{equation}\label{eq:psnr}
\textrm{PSNR}=10\cdot \log_{10}\left({I_{\textrm{max}}^2}/{\textrm{MSE}}\right)\left[\textrm{dB}\right],
\end{equation}

\noindent with MSE representing the Mean Squared Error and the maximum pixel value of the image before compression \(I_{\textrm{max}}\).

\subsection{VVC Configuration}\label{subsec:vvc}

In this study, the VVC reference software \cite{vtmreference} is used and the MSIs were quantized to 10\,bit precision before VVC coding. The concept could be directly transferred to higher precision (e.g. 12, 14, or 16\,bit) if indicated by the application. For VVC intra-only coding of MSI bands, the default VTM intra encoder configuration is used. By using VVC inter-coding for inter-band prediction of MSIs, concepts from  temporal prediction are applied to the task of spectral decorrelation. For encoding the MSI, we adapt the default VTM random access configuration. This configuration uses a Group Of Pictures structure of 32 pictures (GOP-32) and a dyadic hierarchy with bi-prediction for encoding. For adaptation to the MSI data set used here, we drop the first and last pictures on the lowest hierarchy level of the structure, resulting in a GOP-30 structure. By coding the first and last band both as key pictures (highest hierarchy level), this structure can be readily applied to the 31 bands of the MSI. Note that for simplicity, the bands are kept in sequential order here. The same QP was used for all bands but the key pictures, since opposed to pictures of a video sequence, accurate reconstruction is assumed to be of equal importance across all bands. Key pictures are encoded with a QP offset of -3. This configuration will be referred to as "plain" VVC in the following. In an initial experiment, it showed superior compression performance compared to other simple GOP structures. More sophisticated GOP structures adapted to the spectral characteristics of MSIs are subject of further study. 

\subsection{PCA-VVC}\label{subsec:pca_vvc}

PCA-VVC describes a combination of a PCA for spectral and VVC intra for spatial decorrelation. The PCA is applied to the spectral pixel vectors of an MS image, resulting in a concatenation of \(B\) 2D principle component (PC) images, where \(B\) is the number of bands in the MSI. For encoding, only the first $n_\textrm{c}$ PCs representing the highest energy in the data are kept, resulting in a lossy reconstruction.  The PCA basis vectors have to be transmitted together with the PC images for reconstruction on the decoder side. Before transmission, the PCA basis vectors are quantized using a uniform quantizer with a step size of \(\Delta_{v}{=}2^{-13}\).

\subsection{HPCLS}\label{subsec:hpcls}

\begin{figure}[t]
\subfloat[]{
\includegraphics[width=0.5\textwidth]{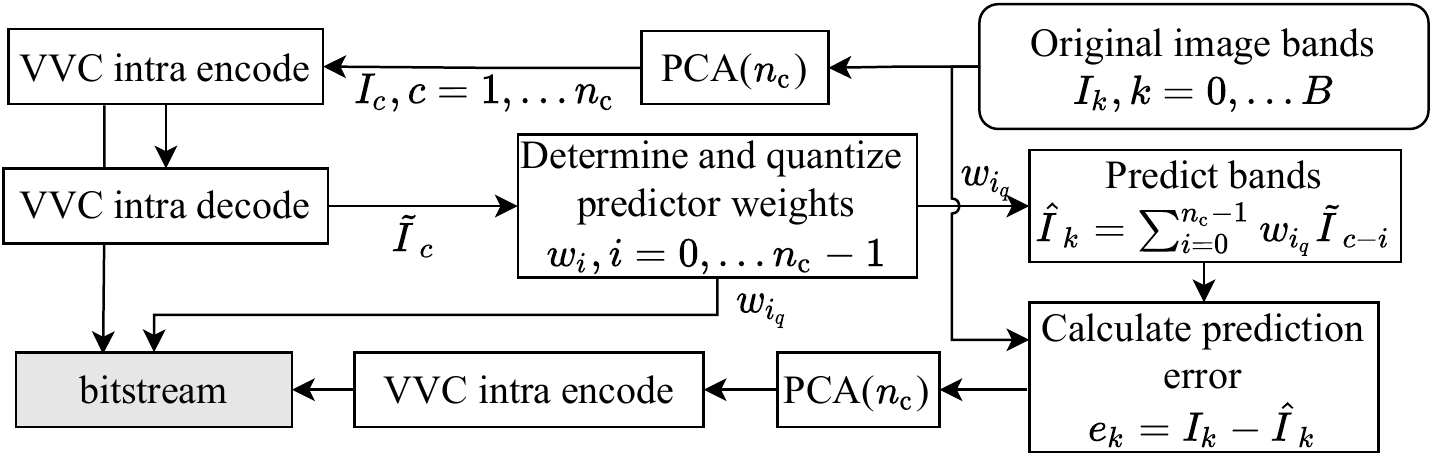}
\label{fig:hpcls}
}
\subfloat[]{
\includegraphics[width=0.5\textwidth]{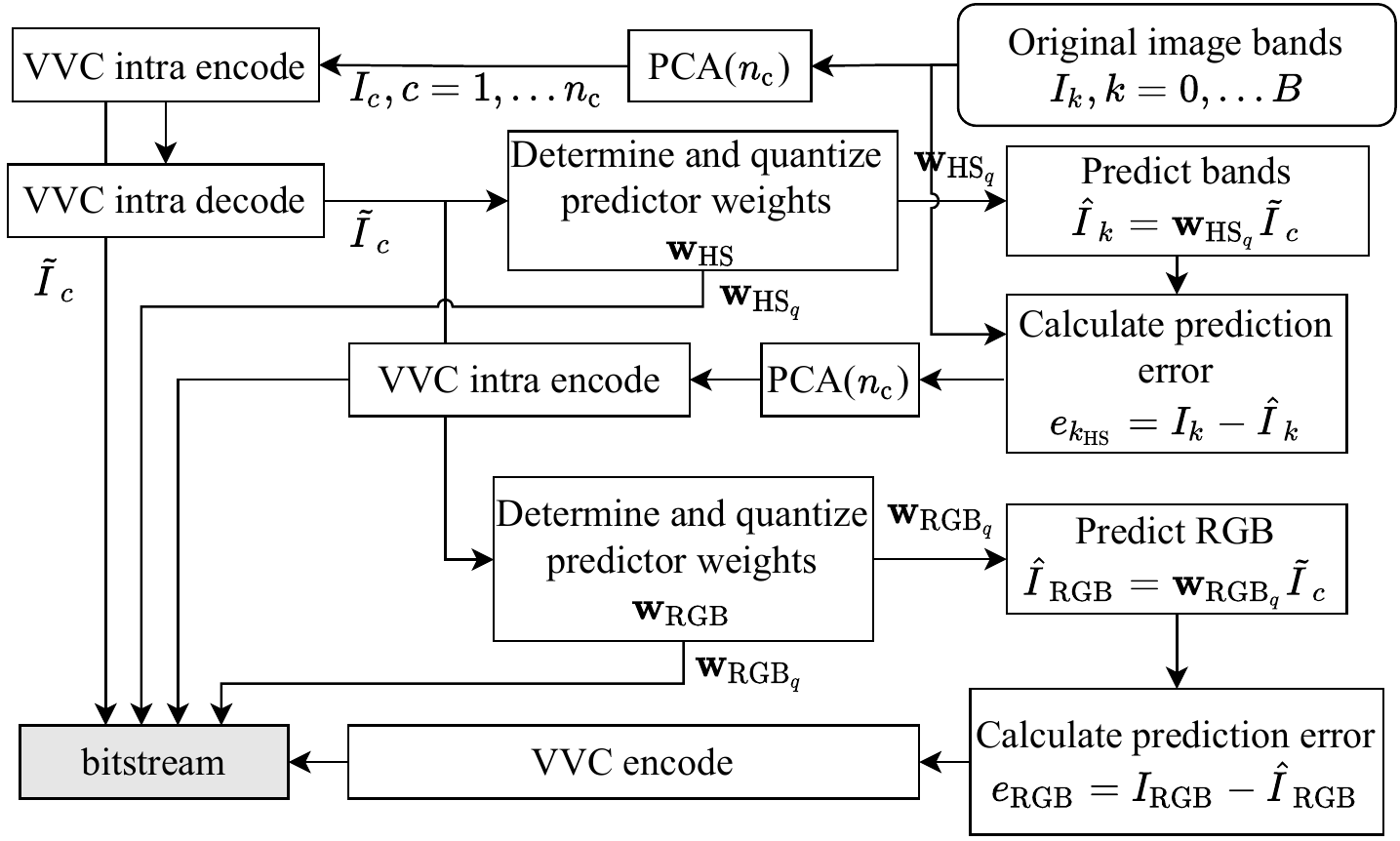}
\label{fig:rgbls}
}
\caption{\protect\subref{fig:hpcls} Block diagram of HPCLS. \protect\subref{fig:rgbls} Block diagram of HPCLS-RGB.}
\label{fig:algorithms}
\end{figure}

The coding scheme of HPCLS is visualized in Fig.~\ref{fig:hpcls}. To obtain the reference bands used for inter-band prediction, a PCA retaining the first \(n_\textrm{c}\) PCs is applied to the MSI as described in section \ref{subsec:pca_vvc}. For this PCA, \(n_\textrm{c}\) will correspond to the number of reference bands \(n_{\textrm{ref}}\). The reference bands are further VVC intra coded. For every spatially non-overlapping square MSI block of size \(S_\textrm{p} \times S_\textrm{p}\) and for every band, predictor weights \(w_i\) are calculated via LS, that predict the MSI from the decoded reference bands \(\tilde{I}_c\). To account for quantization of the \(w_i\) at the decoder side, the prediction error is calculated between the original MSI and the prediction using quantized \(w_i\). The resulting 3D prediction error image is then further compressed by a PCA-VVC. The compressed file comprises the prediction error image and PCA basis vectors, the reference bands and the predictor weights.

\subsection{HPCLS-RGB}

The HPCLS-RGB algorithm is the scalable extension of HPCLS. A second set of weights performs a linear prediction of the RGB preview, which is calculated from the MSI by applying the CIE Standard Observer spectral sensitivities as done in \cite{arad2020ntire}. The resulting prediction error of the RGB preview \(e_{\textrm{RGB}}\) is then directly encoded using VVC in a GOP-2 configuration and added to the bitstream together with the quantized predictor weights \(\mathbf{w}_{\textrm{RGB}_q}\). Fig.~\ref{fig:rgbls} shows a schema of the algorithm.

\section{Experiments}\label{sec:experiments}

The compression algorithms introduced in section \ref{sec:methods} are tested for different parameter configurations on the evaluation data set described in section \ref{sec:dataset}. In order to narrow down the parameter search space, all algorithms were first run in a test set up investigating a bigger variation of parameter settings. In the test set up, the VVC codec was substituted by a 2D DCT for spatial decorrelation, and compressed file sizes were entirely estimated through entropy calculations. The same was done for the predictor weights and PCA basis vectors. Repeating these runs using VVC for a careful selection of parameter settings revealed comparable rate distortion behavior. Parameter preset choices in the following sections were thus based on findings from the test runs.

To obtain the points belonging to one rate-distortion (RD) curve for HPCLS and HPCLS-RGB, the quantization parameter (QP) used for VVC intra-coding of the PCs of the prediction error image is increased by \(5\) within \(\left[5, 50\right]\). RD curves for PCA-VVC are generated using the same QP values for intra-coding of the MSI's PCs. The results are further averaged over the entire 40 image evaluation set to yield the final plots. 

\subsection{Non-scalable Adaptive Compression}

\begin{figure}[t]
\centering{\includegraphics[width=0.5\textwidth]{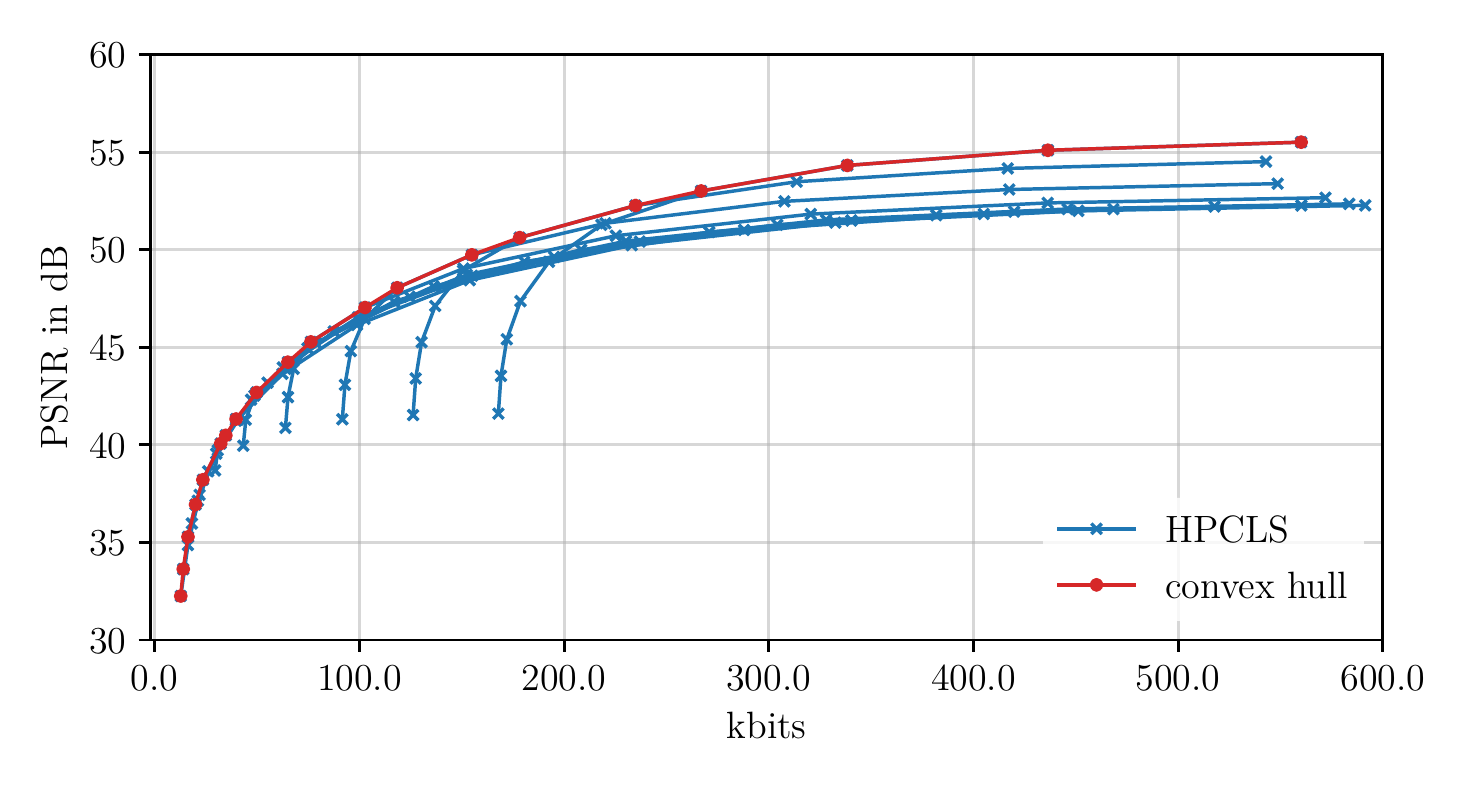}}
\caption{HPCLS for varying \(q_{\textrm{ref}}\) with \(n_{\textrm{ref}}{=}1\) and \(n_{\textrm{c}}{=}3\) fixed. Lower \(q_{\textrm{ref}}\) shift the RD curve towards the to right. The convex hull is established by the best-performing rate points.}
\label{fig:hpcls_example}
\end{figure}

HPCLS is run for \(n_{\textrm{ref}}{\in}\{1, 2, 3\}\), varying QP for compression of the reference bands \(q_{\textrm{ref}}{\in}\left[5, 50\right]\) and varying number of PCs for compression of the prediction error \(n_{\textrm{c}}\in \left[1, 6\right]\). PCA-VVC is run for different \(n_{\textrm{c}}{\in}\left[1, 10\right]\). The prediction block size is preset to \(S_p{=}64\). The PCA basis vectors \(v\) and the HPCLS and HPCLS-RGB predictor weights are quantized with a step size of \(\Delta_{v}{=}2^{-13}\) and \(\Delta_{w}{=}2^{-12}\), respectively\footnote{As of now, the \(w_i\) and PCA basis vectors are fixed length coded, and coding cost could very likely be reduced, e.g. by an entropy coding stage.}. Figure \ref{fig:hpcls_example} shows the RD plots for an HPCLS example run where only \(q_{\textrm{ref}}\) was varied, plotted in blue. In order to investigate the best performance a encoder might achieve, the convex hull of RD points generated by varying the parameter settings of each scheme is reported. The process is illustrated in Fig.~4. The design of the parameter adaptation is subject of further study. In this manner, the convex hull RD curves of all investigated PCA-VVC and HPCLS configurations are plotted against the curve belonging to ``plain'' VVC coding of the entire MSI in Fig.~\ref{fig:non_scalable}. Besides the modified GOP structures, the VTM configurations are used as-is.

\subsection{Scalable Adaptive Compression}

\begin{figure}[t]
\centering{\includegraphics[width=0.5\textwidth]{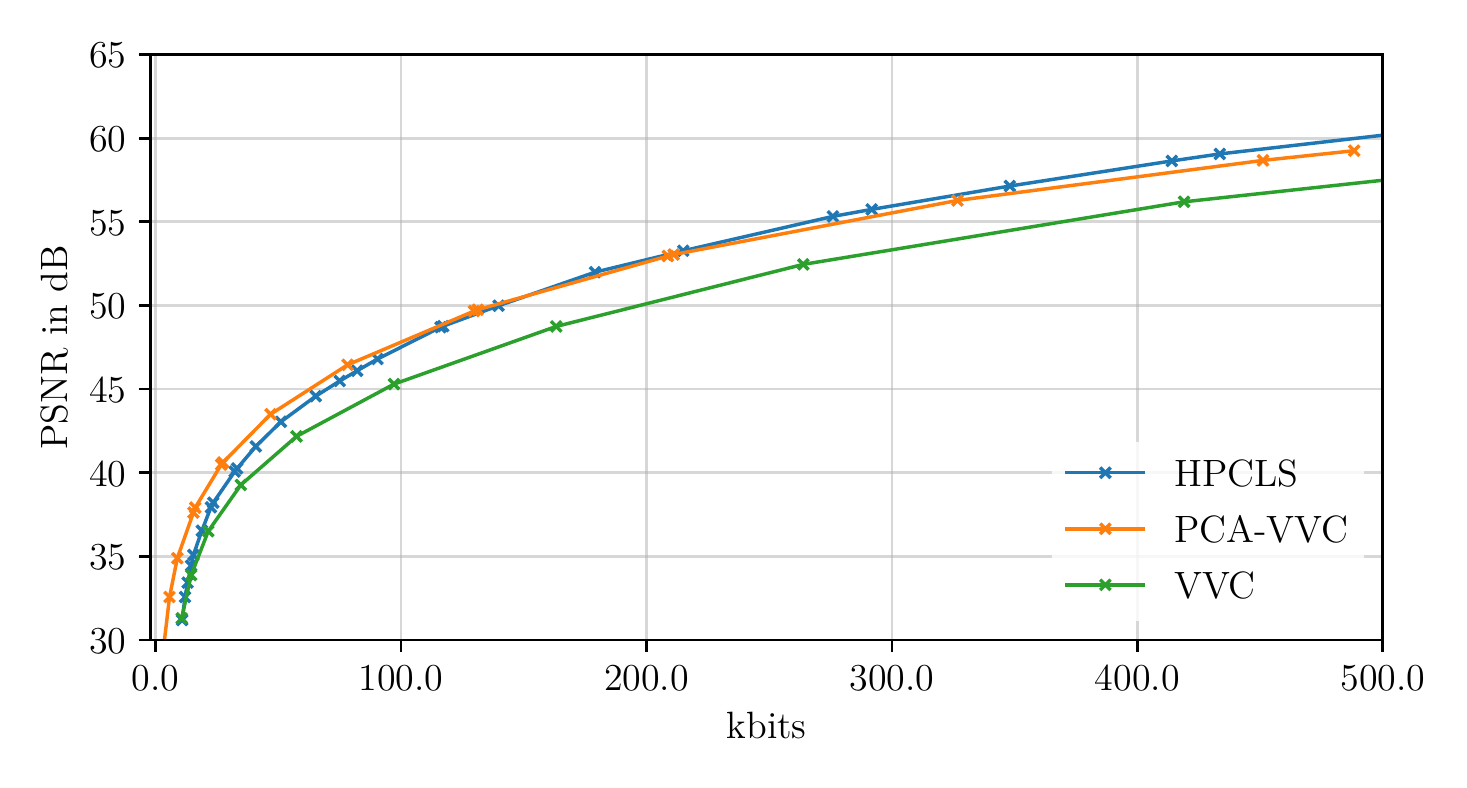}}
\caption{Comparison of HPCLS, PCA-VVC and ``plain'' VVC.}
\label{fig:non_scalable}
\subfloat{
\includegraphics[width=0.5\textwidth]{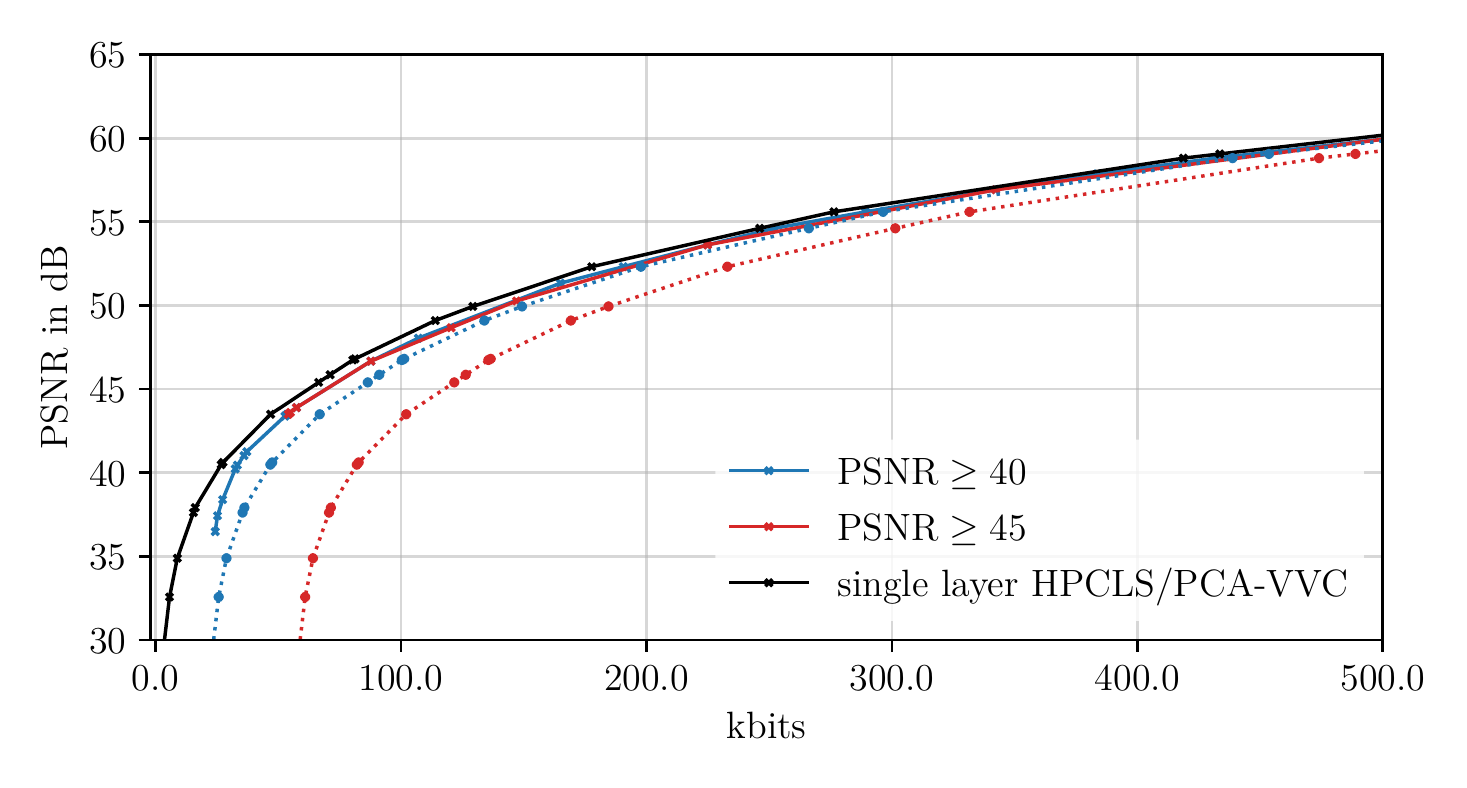}
\label{fig:scalable1}
}
\subfloat{
\includegraphics[width=0.5\textwidth]{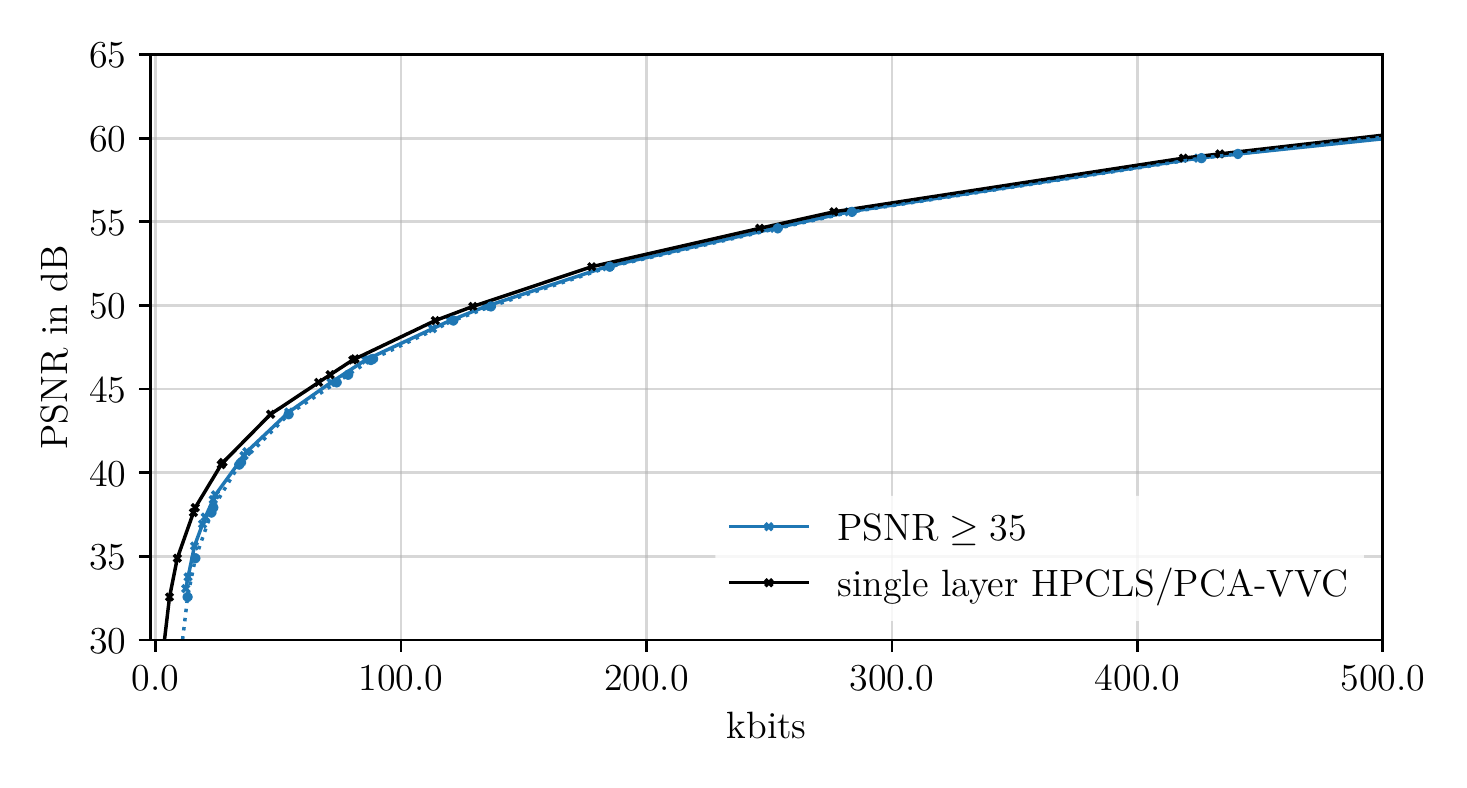}
\label{fig:scalable2}
}
\caption{Comparison of simulcast HPCLS/PCA-VVC (dotted) and HPCLS-RGB (solid) using adaptive and scalable coding structures for three different qualities of the RGB preview (35, 40 and 45\,dB PSNR), plotted against single layer HPCLS/PCA-VVC, that does not provide a preview.}
\label{fig:scalable}
\end{figure}

The HPCLS-RGB algorithm is run for the same parameter settings as done for HPCLS while additionally varying the QP used to encode the RGB preview. By this, the performance of compression method can be evaluated for different RGB preview qualities. As done for HPCLS, the predictor weight quantization step size and the prediction block size are preset to \(\Delta_w{=}2^{-12}\) and \(S_p{=}64\), respectively. For HPCLS-RGB to be a feasible option for scalable compression, it should provide superior RD performance when compared to simply adding an RGB preview to the compressed files produced by the non-scalable approaches. The latter technique is referred to as ``simulcast'' in the following. Fig.~\ref{fig:scalable} compares the scalable HPCLS-RGB to the simulcast version of HPCLS and PCA-VVC for different preview qualities measured in PSNR. The curve of non-scalable coding by HPCLS/PCA-VVC is added to illustrate the coding cost for the scalability feature. Note that the convex hull of HPCLS and PCA-VVC have been combined to a single best performing HPCLS/PCA-VVC curve here.

\section{Discussion}\label{sec:discussion}

The results plotted in Fig.~\ref{fig:non_scalable} reveal that the proposed adaptive HPCLS algorithm outperforms the PCA-VVC approach for reconstruction qualities above 51\,dB PSNR on the evaluated image set. Note that this crossing point might be significantly shifted to lower PSNRs by developing an entropy encoding scheme for the HPCLS predictor weights. This makes HPCLS a viable option, since MSIs are usually used at very high qualities. Both, PCA-VVC and HPCLS outperform ``plain'' VVC coding for reconstruction qualities between 30 and 60\,dB. At 50\,dB, both PCA-VVC and HPCLS achieve a compression ratio of roughly 1:100.

The compression performance of the proposed scalable HPCLS-RGB compared to the non-scalable simulcast  versions of HPCLS and PCA-VVC depends on the desired quality of the RGB preview and the reconstructed MSI. As can be seen from Fig.~\ref{fig:scalable}, HPCLS-RGB outperforms the simulcast configuration significantly.

\section{Conclusions}\label{sec:conclusion}

In this work, the VVC codec was applied to the task of MSI compression using adaptive and scalable coding structures. VVC inter prediction was found to be outperformed by the PCA for spectral decorrelation of the MSIs. With HPCLS, a novel adaptive MSI compression algorithm combining concepts from transform and predictive coding was introduced, that outperforms the combination of PCA for spectral and VVC  intra-coding for spatial compression for reconstruction qualities above 51\,dB PSNR. Further, an adaptive salable approach was developed, that provides a separately decodable RGB preview of the MSI's contents. The RD performance of the scalable approach is superior to simply adding the RGB preview to the compressed files produced by the non-scalable approaches.

In future work on scalable MSI compression, the design of the preview may be revisited in order to achieve better exploitation of its spectral information for prediction.
This could be achieved e.g. by replacing the RGB preview with the output of a colour transform which is more optimized for the compression task. For HPCLS-RGB and HPCLS, the linear predictor might be substituted by a small multilayer perceptron or convolutional network, that is trained and transmitted for each individual image.
The proposed compression algorithms will further be evaluated on other MS data sets.

\Section{References}
\bibliographystyle{IEEEtran}
\bibliography{literatur}

\end{document}